\documentclass[aps,twocolumn]{revtex4-2}
\usepackage{amsmath, amssymb, amsfonts}
\usepackage{mathrsfs}
\usepackage{graphicx}
\usepackage{bm}
\usepackage{multirow}
\usepackage{float}
\usepackage{xpatch}

\newcommand{\expect}[2][]{\operatorname{\textnormal{\textbf{E}}}_{#1}\left[#2\right]}

\newcommand{\mat}[1]{\bm{#1}}

\newcommand{\params}{\ensuremath{\mat{\theta}}}

\newcommand{\data}{\ensuremath{\mat{y}}}
\newcommand{\Rnot}{\mathcal{R}}
\newcommand{\osize}{\mathcal{O}}
\newcommand{\imax}{\mathcal{P}}
\newcommand{\grate}{\mathcal{G}}
\newcommand{\tpeak}{\mathcal{T}}

\begin{document}

\title{Accurately summarizing an outbreak using epidemiological models takes time}

\author{B. K. M. Case\textsuperscript{1,3}}
\author{Jean-Gabriel Young\textsuperscript{1,2,3}}
\author{Laurent H\'ebert-Dufresne\textsuperscript{1,3}}
\email{Email: laurent.hebert-dufresne@uvm.edu}

\affiliation{\textsuperscript{1}Department of Computer Science, University of Vermont, Burlington, 05405, USA}
\affiliation{\textsuperscript{2}Department of Mathematics \& Statistics, University of Vermont, Burlington, 05405, USA}
\affiliation{\textsuperscript{3}Vermont Complex Systems Center, University of Vermont, Burlington, 05405, USA}

\date{\today}

\begin{abstract}
Recent outbreaks of monkeypox and Ebola, and worrying waves of COVID-19, influenza and respiratory syncytial virus, have all led to a sharp increase in the use of epidemiological models to estimate key epidemiological parameters.
The feasibility of this estimation task is known as the \emph{practical identifiability} (PI) problem.
Here, we investigate the PI of eight commonly reported statistics of the classic Susceptible-Infectious-Recovered model using a new measure that shows how much a researcher can expect to learn in a model-based Bayesian analysis of prevalence data. 
Our findings show that the basic reproductive number and final outbreak size are often poorly identified, with learning exceeding that of individual model parameters only in the early stages of an outbreak.
The peak intensity, peak timing, and initial growth rate are better identified, being in expectation over 20 times more probable having seen the data
by the time the underlying outbreak peaks.
We then test PI for a variety of true parameter combinations, and find that PI is especially problematic in slow-growing or less-severe outbreaks.
These results add to the growing body of literature questioning the reliability of
inferences from epidemiological models when limited data are available.
\end{abstract}

\maketitle

% \section*{Introduction}

Incredible efforts have been made in recent years to apply epidemiological models to the empirical data borne out of the COVID-19 pandemic.
The LitCovid aggregator currently contains over 3,000 papers on ``epidemic forecasting'' and ``modelling and estimating'' trends of COVID-19 spread \cite{chen2021litcovid}. 
We are seeing similar waves of models and forecasts for recent outbreaks of monkeypox, Ebola, influenza and respiratory syncytial virus. 
However, the enormous variability in model predictions, even among works using the same model and similar data, erodes confidence when interpreting these efforts for policy decisions \cite{Roda2020}. It is clear that uncertainty remains about what we can expect to learn from models, and when.

Disease models tackle the difficult challenge of describing complex epidemic processes by relating mechanistic processes to population level observations such as daily reported cases.
Identifying combinations of parameters which plausibly replicate observed data can help summarize the epidemic dynamics. 
Common statistics include the basic reproductive number $\Rnot$, the average number of new cases someone will cause in an entirely susceptible population, and the outbreak size, the fraction of the population who will eventually have had the disease.
Because these indicators are the product of interacting social and biological phenomena, they are never available through direct observation.
Fitting epidemiological models to data is one of the best options for estimating these important quantities \cite{Wu2020}.

The classic Susceptible-Infectious-Recovered (SIR) model accounts for a minimal number of critical mechanisms of disease spread. Infectious individuals infect susceptible individuals at a rate $\beta$ and recover at a rate $\alpha$. These mechanisms can be tracked through time by a set of ordinary differential equations:
\begin{equation*}
\frac{d}{dt}S = -\beta SI \; , \quad \frac{d}{dt}I = \beta SI - \alpha I \; , \quad \textrm{and} \quad \frac{d}{dt}R = \alpha I \; .
\end{equation*}
It is common to consider $S$, $I$ and $R$ as a fraction of the population in a given state such that $S+I+R=1$ at all time.
The initial state of the population  might not be known---especially the susceptible pool $S_0 \equiv S(t=0)$. 
Focusing on the second equations, we can see that the epidemic will grow exponentially at a rate $\beta S_0 - \alpha$ for initial small values of $I$, making it clear there will be large uncertainty in the value of individual parameters \cite{melikechiLimitsEpidemicPrediction2022}.
Conversely, when $I$ becomes small after the peak, the infectious population eventually decays exponentially at a rate $\alpha$.
Observations of $I$ will therefore provide information about different parameters, or combinations thereof, at different points of an outbreak.
However, how this information accumulates over time and how it allows us to identify key summary statistics is more complicated.

\begin{figure*}[t]
\centering
    \includegraphics[width=\linewidth]{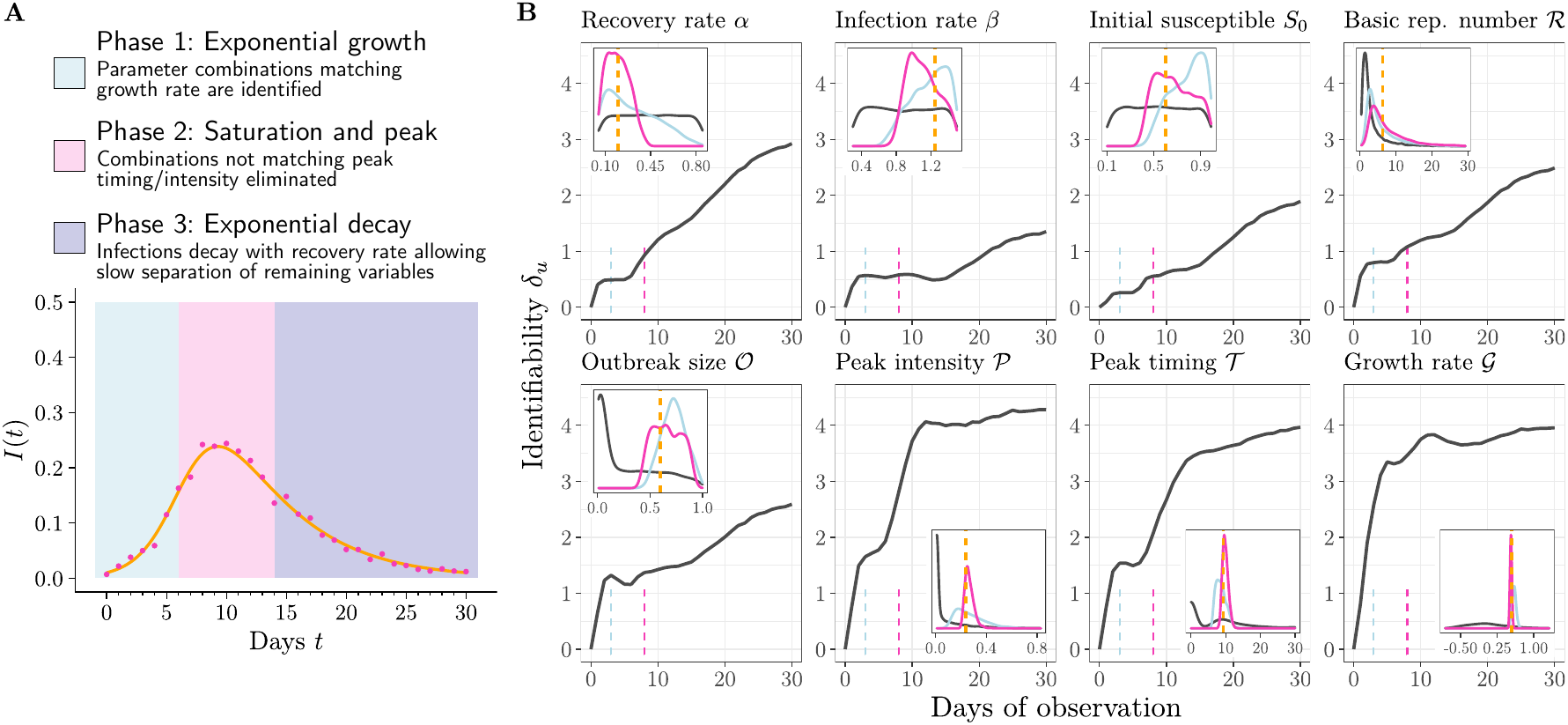}
    \caption{
    \small
    Practical identifiability (PI) of epidemiological summary statistics over time.
    (\textbf{A}) Unknown deterministic SIR process based on true parameters $\params^*$ (orange line), and single realization of observed data $\data \sim P (\data \mid\params^*)$ (pink dots). 
    (\textbf{B}) 
    Main panels show PI according to $\delta_u$ over an increasing observation window assuming daily observations.
Insets give an example of how $\delta_u$ is interpreted, showing $P(u \mid \data)$ and $P(u)$ for the single realization of $\data$ from (A), observed up to $T = 3$ (blue) and $T = 8$ (pink). The dashed orange line is the
true value to be estimated.
True parameters are $\alpha^*=0.2$, $\beta^*=1.25$, and $S_0^*=0.6$, with $I_0=0.01$ assumed known. Prior beliefs are $\alpha\sim U(0.05, 0.85)$, $\beta\sim U(0.3, 1.5)$, $S_0\sim U(0.1, 0.99)$.
}
    \label{fig:gain-over-time}
\end{figure*}

The widespread application of models built 
on the SIR backbone has led several authors to caution that the reliability of predictions
can be sensitive to available data \cite{castroTurningPointEnd2020, melikechiLimitsEpidemicPrediction2022}, and even more so for common extensions such as the SEIR model \cite{tuncerStructuralPracticalIdentifiability2018, Roda2020}.
The question of whether parameters estimated from data are reliable, i.e. close to some hypothetical true parameters $\params^* = (\alpha^*, \beta^*, S_0^*)$ which generated the data, is termed the \emph{practical identifiability} (PI) problem.
Here we 
use
a new measure which allows us to directly measure our ability to learn various epidemiological quantities.
If $u=f(\params)$ is an unknown variable to be estimated, our pseudo-Bayesian interpretation of the identifiability of $u$ is the expected logarithm of the ratio between posterior and prior probabilities, evaluated at~$u^*$:
\begin{align}
    \delta_{u}(\params^*) = \expect[\data\mid\params^*]{\log P(u^*\mid \data)  - \log P(u^*)}
    \label{eq:rmd-bayes}
\end{align}
where $\data\mid\params^*$ are noisy observations of the epidemiological variable, e.g., daily case counts, and where the expectation is taken over realizations of the observation process.
This measure reflects the magnitude of information a researcher can expect to gain when fitting a model to data, while allowing the effect of particular values of $\params^*$ to be studied.
\eqref{eq:rmd-bayes} does not require computationally expensive Bayesian inference methods to compute -- a simple Monte Carlo procedure for estimating \eqref{eq:rmd-bayes} is outlined in the SI Text.

\section*{Results}

Figure \ref{fig:gain-over-time} shows the PI of the SIR model parameters, as well as five summary variables which are commonly calculated in terms of $\params$ (\textbf{see Table \ref{tab:epi-vars} for mathematical definitions}), for a typical parametrization $\params^*$ of the model.
Observations were distributed with relatively little noise, to better study PI inherent to the SIR model itself.
$\delta_u$ is computed daily for these eight variables using observations for the first 30 days.

The rate of learning for all variables is uneven over time, with each reaching plateaus of varying length before the peak. 
The infection rate $\beta$ is the worst identified.
Gaining information on $\alpha$ appears easier than $\beta$ and $S_0$ 
and even exceeds learning for $\Rnot$ and $\osize$ after around $T=20$ days of observation. 
PI of the peak intensity, peak timing, and growth rate increase more rapidly at first, with the learning of the growth rate happening particularly fast.
The true growth rate is over 25 times more probable having seeing the data after only 5 days of observation.

To test the sensitivity of these findings to $\params^*$, we computed $\delta_u$ over a grid of values for $\beta^*$ and $S_0^*$ (Figure \ref{fig:gain-var-true}). Since slower growing outbreaks will naturally contain less information per day \cite{capaldiParameterEstimationUncertainty2012}, information gain was calculated using observations up until the first day after the epidemic peak. 
The outbreak size of the true epidemic was the most correlated with learning of the five summary variables, followed by growth rate.

\section*{Discussion}

\begin{figure*}[t]
\centering
\includegraphics[width=\linewidth]{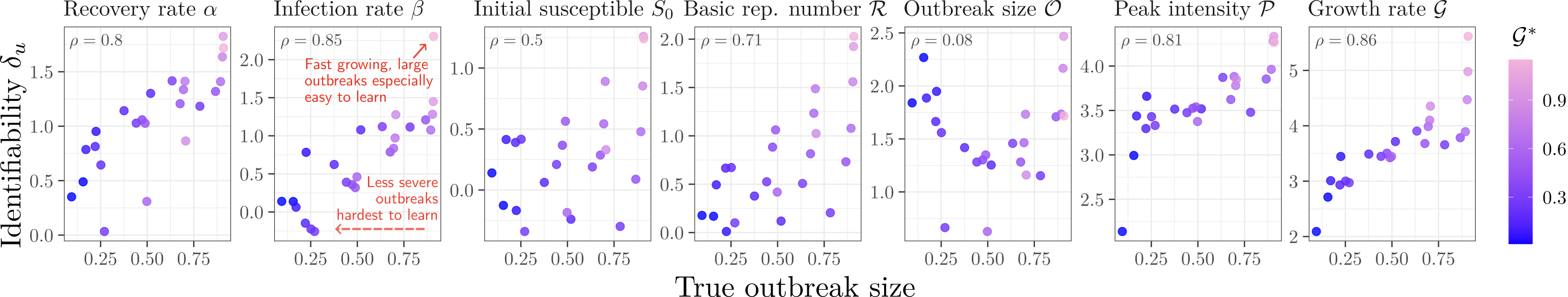}
\caption{\small Change in practical identifiability $\delta_u$ when the true parameters $\params^*$ are varied. $\delta_u$  is calculated using daily observations up to the first day after the true (unobserved) outbreak has peaked. 
True parameters tested were all combinations of $\beta^*=0.3, 0.5, \dots, 1.5$ and $S_0^*=0.1, 0.3, \dots, 0.9$ with $\alpha^*=0.2$ fixed. 
Pearson correlation between $\delta_u$ and true outbreak size is given in corners of each panel.
Priors are the same as in Figure \ref{fig:gain-over-time}.
}
\label{fig:gain-var-true}
\end{figure*}

The analysis presented here makes it clear that some epidemiological variables are easier to estimate through model dynamics than others, and emphasizes that most epidemiological summary statistics should be interpreted with caution when data are limited. 
Taken together, the rate of learning for all the variables suggests that learning takes place in three general phases. 
In phase 1, plausible parameter combinations quickly concentrate along the surface $\{\params : \beta S_0 - \alpha = \grate^*\}$, as infections increase exponentially with the initial growth rate. This explains the sharp but modest gain in information of all variables except for $\grate$ during this phase. In phase 2, infections begin to saturate and parameter combinations matching the true peak intensity and timing become more plausible. However, for $\beta$ especially, saturating case counts do little to further restrict the plausible parameter surface from phase 1. Finally, phase 3 is characterized by gradual information gain for the remaining variables. Since infections are slowly declining with $\alpha$ during this phase, this growth is explained by $\alpha^*$ gradually being identified, which propagates to allow some remaining combinations on the plausible surface to be eliminated.

Parameters describing the mechanisms of the model---$\beta$, $\alpha$ and $S_0$---take a particularly long time to learn on account of quickly reaching a plateau at low values of $\delta_u$.
As a result, the SIR model is more effective at forecasting short term statistics of the dynamics such as peak timing and intensity, than it is at estimating mechanisms. This result shows how difficult it is to estimate parameters from early data in the hope of forecasting the impacts of mechanistic interventions such as reducing $\beta$ with preventive measures or increasing $\alpha$ with treatment \cite{Barnett2023}.

Learning was nearly as difficult for the statistics $\Rnot$ and 
$\osize$ as for the individual model parameters, despite the fact that optimistically, these transformations would combine the information of each parameter they depend on. 
The failure of these statistics to resolve closely exchangeable parameter combinations limits their reliability for succinctly describing an epidemic.
In contrast, the initial growth rate resolves such combinations to give rapid shrinkage to the correct value, despite encoding similar information as $\Rnot$ about disease dynamics \cite{Bettencourt2008}. This suggests growth rates are a more reliable ``first look'' at an outbreak when using prevalence data under the SIR model.

When varying the true values $\params^*$, see Figure \ref{fig:gain-var-true}, we find 
that less-severe outbreaks are generally harder to learn, despite having more daily observations available before their peak.
The initial susceptible population $S_0$ appears the most poorly identified across values of $\params^*$ by the peak, and the expected posterior shrinkage is even slightly negative for 25\% of the tested values.
An interesting implication for control measures is that the more we reduce the severity of true infection dynamics, the harder it will be to accurately estimate the impacts of interventions. 
Further, the mode of intervention matters: variability along the y-axis in Figure \ref{fig:gain-var-true} for similar values of $\osize^*$ shows lowering $S_0^*$ impacts learning differently than a reduction in $\beta^*$.

Previous investigations into the PI of the SIR model have mainly focused on the PI of $\alpha$ and $\beta$ under the simplified model where $S_0 \approx 1$ is known. 
These works generally agree that PI of both $\alpha$ and $\beta$ is limited during 
phase 1 \cite{melikechiLimitsEpidemicPrediction2022}, but that the majority of information available has been learned by the time the disease has peaked \cite{tuncerStructuralPracticalIdentifiability2018, piazzolaNoteToolsPrediction2021}. 
Most comparably to the observational design in Figure \ref{fig:gain-over-time}, Capaldi et al. (2012) considered the asymptotic variance of $\hat{\beta}$ and $\hat{\alpha}$ over an increasing timespan, and found the variance of both estimators decreased rapidly and smoothly just before and after the peak, respectively \cite{capaldiParameterEstimationUncertainty2012}.
In contrast, the delayed rate of learning of these parameters in Figure \ref{fig:gain-over-time} paints a more pessimistic picture of PI when exact likelihoods and prior context is taken into account. This finding supports the idea that previous PI results based on sensitivity equations underestimate uncertainty, particularly during the early stages of an outbreak when the likelihood surface is highly nonlinear \cite{joshiExploitingBootstrapMethod2006, hinesDeterminationParameterIdentifiability2014}.

The Bayesian nature of our method of assessing PI means that estimates of model parameters and any variables which depend on them are sensitive to prior beliefs. In this report, our choice of uniform priors represents modest assumptions about an emerging pathogen: \emph{a priori}, just over 50\% of scenarios result in an outbreak (i.e. have $\beta S_0 / \alpha > 1$), and outbreaks range from modest to highly severe (70\% of individuals infected at peak).
However, for many pathogens, more informative prior information is frequently available, for example on the recovery rate of a disease \cite{Cevik2021}. 
Relative to more realistic settings for $P(\params)$, this may mean $\alpha$ is more difficult to gain information about than $\beta$ and $S_0$.
Further, our choice of priors shows that initial shrinkage in the likelihood surface can just as readily be explained by common-sense bounds on the model parameters. 
In this sense, not taking prior assumptions into account when calculating PI arguably over-reports learning.

While we have considered only noisy observation of the current infectious population, real data may also come in the form of daily new infections or cumulative case counts, and may suffer from lags in reporting or preferential sampling \cite{chiuUsingTestPositivity2021, wuSubstantialUnderestimationSARSCoV22020}. Learning epidemiological variables from such data will have their own distinct challenges \cite{tuncerStructuralPracticalIdentifiability2018}. 
PI of the SIR model should also be assessed with hierarchical models incorporating data from multiple sources, such as hospitalizations and isolated clinical experiments \cite{deangelisFourKeyChallenges2015a}. Yet, our work shows that even in its simplest form, learning parameters and statistics of SIR dynamics takes time, limiting which inferences, forecasts, and control policies can be made from early epidemic data.

\section*{Methods}

\subsection*{Data availability} 
\noindent Materials necessary to reproduce this analysis are available online at \url{github.com/brendandaisy/epi-summaries-over-time}.

\subsection*{Observation model}

\noindent Infectious individuals are assumed to be independently tested at a fixed rate $\eta$ at integral timepoints $t=1,\dots,T$, giving a likelihood $y_t \sim \textnormal{Poisson}(\eta I(t; \params^*))$, where $I(t; \params^*))$ are the infectious dynamics parameterized by unknown values $\params^*$. $\eta=1000$ is assumed known throughout.

\section*{Acknowledgements}

\noindent BC, JGY and LHD acknowledge support from the National Institutes of Health 1P20 GM125498-01 Centers of Biomedical Research Excellence Award. BC is also supported as a Fellow of the National Science Foundation under NRT award DGE-1735316, and LHD by the National Science Foundation award EPS-2019470.

\appendix
\section{Supplemental Methods}

\subsection*{Likelihood-based estimation of dynamical systems}

While the methods considered here can be applied to any statistical process for which a likelihood exists, we are interested in processes of the form
\begin{gather}
    y_i = g(\mat{x}(t_i), \mat{\sigma}) \label{eq:obs-proc}\\
    \mat{\dot{x}}(t) = h(\mat{x}(t), \mat{\tau}) \label{eq:lat-proc}
\end{gather}
where $\data=(y_1, \dots, y_n)$ are observations at discrete timepoints $t_1,\dots,t_n$, and $\mat{\sigma}$, $\mat{\tau}$ are parameters that are assumed known or are to be estimated. We refer to $h$ as the \emph{latent process} and $g$ as the \emph{observation process}. We are interested in our ability to estimate a set of unknown parameters $\params^* \subseteq (\mat{\sigma}^*, \mat{\tau}^*, \mat{x}(0)^*)$. 

Given $\params$, Eqs. (\ref{eq:obs-proc}-\ref{eq:lat-proc}) form a probability distribution $P(\data\mid\params)$ referred to as the \emph{likelihood}. In the frequentist paradigm, an estimator for $\params^*$ can be obtained by maximizing $P(\data\mid\params)$, 
\begin{equation*}
    \hat{\params}_{\textnormal{MLE}} = \textnormal{argmax}_{\params} P(\data\mid\params)
\end{equation*}

A popular way to assess issues of practical identifiability is through the variance-covariance matrix of $\hat{\params}_{\textnormal{MLE}}$, which can show marginal uncertainty in individual parameter estimators and correlations between pairs of estimators. 
The Cramer-Rao bound states that in the limit of decreasing observation uncertainty (i.e. as the amount or precision of data increases), the variance of an unbiased estimator converges, given certain regularity conditions, to the inverse of the Fisher Information Matrix $\mathcal{I}(\params^*)$, where
\begin{equation}
    \left[\mathcal{I}(\params)\right]_{ij} = -\expect[y\mid\params]{\frac{\partial^2}{\partial\theta_i\partial\theta_j} \log P(y\mid\params)}.
    \label{eq:info-mat}
\end{equation}

This bound can underestimate variance when measurement noise is not infinitesimal \cite{ramanDelineatingParameterUnidentifiabilities2017, hinesDeterminationParameterIdentifiability2014}, leading some to question its applicability even for simple nonlinear models \cite{joshiExploitingBootstrapMethod2006, krauschMonteCarloSimulations2019}.
An alternative is to estimate the distribution of $\hat{\params}_{\textnormal{MLE}}$ using Monte Carlo simulation, by sampling possible data sets $\data^{(1)}, \data^{(2)}, \dots$ from $P(\data\mid\params^*)$ and finding the maximum of each likelihood $P(\data^{(j)}\mid\params)$ using an optimization algorithm such as gradient descent. 
The resulting samples $\hat{\params}_{\textnormal{MLE}}^{(j)}$ can then be inspected graphically or used to estimate the covariance matrix.
This method has the convenience of also working with estimates of transformations of the model parameters, without the need for further approximation \cite{Chowell2004}.

\subsection*{Proposed method of assessing practical identifiability}

While using Monte Carlo estimation of $\textnormal{Var}(\hat{\params}_{\textnormal{MLE}})$ to assess PI can alleviate the underestimation issues when using the Information Matrix, the use of optimization to obtain a sample of the estimator can lead to dependence on initial conditions or other hyperparameters of the optimization method used \cite{lamPracticalIdentifiabilityParametrised2022}.

We instead take a Bayesian perspective. From the main text, we have for a variable of interest $u=f(\params)$, $
    \delta_{u}(\params^*) = \expect[\data\mid\params^*]{\log P(u^*\mid \data)} - \log P(u^*),
$ which gives the average amount, over possible future outbreaks $P(\data\mid\params^*)$, a researcher can expect to learn about the true quantity $u^*$ in a Bayesian analysis. A value of $\delta_u = c$ corresponds roughly to an expected gain in posterior probability $c$ orders of magnitude greater than the prior.

\eqref{eq:rmd-bayes} can be rewritten by applying Bayes' rule, $P(u^*\mid\data)/P(u^*)=P(\data\mid u^*)/P(\data)$, where the margin $P(\data \mid u^*)$ is equal to $\int P(\data \mid \params) P(\params\mid u^*) d\params$ and $P(\params\mid u^*)$ is the distribution of the epidemiological parameters compatible with a fixed variable of interest $u^*$---we give details below.
This leads to
\begin{equation}
    \delta_u(\params^*) = \expect[\data\mid\params^*]{\log \frac{P(\data \mid u^*)}{P(\data)}}.
\end{equation}
We approximate $\delta_u(\params^*)$ in a numerically stable way by generating $M$ paired Monte Carlo samples from $P(\params\mid u^*)$ and $P(\params)$, and reusing these samples to obtain $M$ samples from $P(\data\mid u^*)$ and $P(\data)$ for each $\data\sim P(\data\mid\params^*)$, leading to
\begin{align*}
    \delta_u(\params^*)\approx \frac{1}{N}\sum_{i=1}^N \Big[ \log \frac{1}{M} \sum_{j=1}^M& P(\data^{(i)}\mid \Tilde{\params}^{(j)})  \\
    &-\log \frac{1}{M} \sum_{j=1}^M P(\data^{(i)}\mid\params^{(j)})\Big]
\end{align*}
where $\Tilde{\params}^{(j)}\sim P(\params\mid u^*)$, $\params^{(j)}\sim P(\params)$, and $\data^{(i)}\sim P(\data\mid\params^*)$.
$N=3000$ and $M=60,000$ was used for all computations in this work. 

\subsubsection*{Accuracy of Monte Carlo estimation of $\delta_u$}

\begin{figure}
    \centering
    \includegraphics[width=\linewidth]{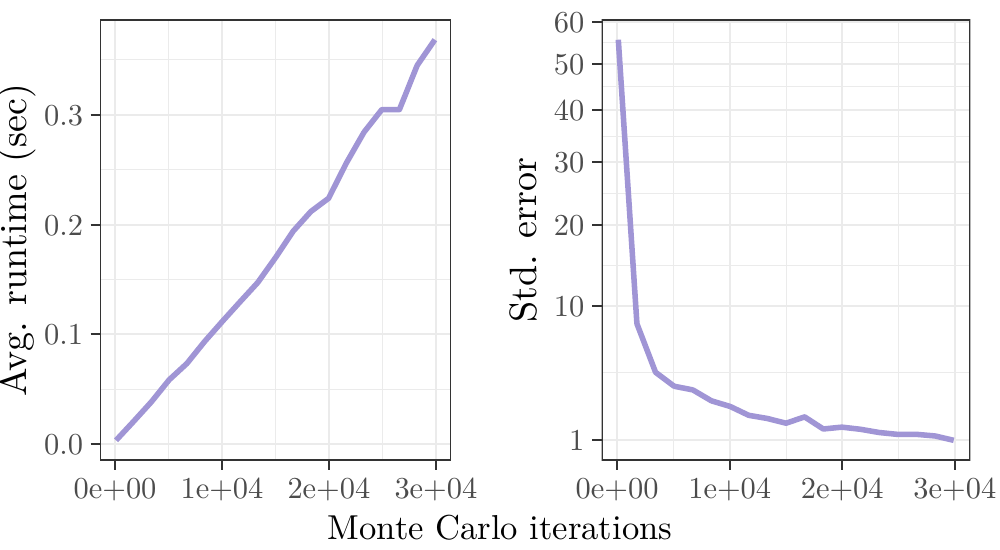}
    \caption{Speed and accuracy of approximating the log marginal likelihood $P(\data)$ using $M$ Monte Carlo simulations. The results are averaged over 100 repetitions of the sampling process for each $M$.}
    \label{fig:marg-lik-std-err}
\end{figure}

The marginal likelihood $P(\data)$ is notorious for being inefficient to estimate via Monte Carlo methods. To test our choice of $M$ was large enough while still within a reasonable computational budget, we repeated calculations of $\log P(\data)$ for increasing values of $M$, where $\data\sim P(\data\mid\params^*)$ was sampled with 60 observations (every half day). $\params^*$ and $P(\params)$ were the same as in Figure 1 of the main text. We concluded that even with 60 observations, which gives a likelihood more sharp than the maximum 30 observations used in the main text, a choice of $M>30,000$ was sufficient to give a standard error less than 1, or less than 0.5\% of the magnitude of $\log P(\data)$.
The runtime and standard errors from 100 independent computations of $P(\data)$ are shown as a function of $M$ in Figure \ref{fig:marg-lik-std-err}.

\subsection*{Practical identifiability for a function of model parameters}

\begin{table}\centering
\caption{Definitions of epidemiological summary statistics}
\begin{tabular}{lcr}
Name & Symbol & Formula \\
\hline
Reproductive number & $\Rnot$ & $\beta/\alpha$\\
%number &&\\
Outbreak size & $\osize$ & $1 - R(0) - S_0\exp(-\Rnot \osize)$\textsuperscript{*}\\
%Peak intensity & $\imax$ & $I_0 + S_0 -$ \\
%&& $\Rnot^{-1}\left(\log S_0 - 1 - \log \Rnot\right)$\\
Peak intensity & $\imax$ & $I_0 + S_0 + \left[1-\log \left(S_0/\Rnot\right)\right]/\Rnot$\\
Peak timing & $\tpeak$ & Unknown\\
Growth rate & $\grate$ & $\beta S_0 - \alpha$ \\
\hline
\textsuperscript{*}Implicit equation &&
\end{tabular}
\label{tab:epi-vars}
\end{table}

The distribution function $P(\params\mid u^*)$ will generally not be available in closed form even when $P(\params)$ is.

\begin{figure*}
    \centering
    \includegraphics[width=.65\linewidth]{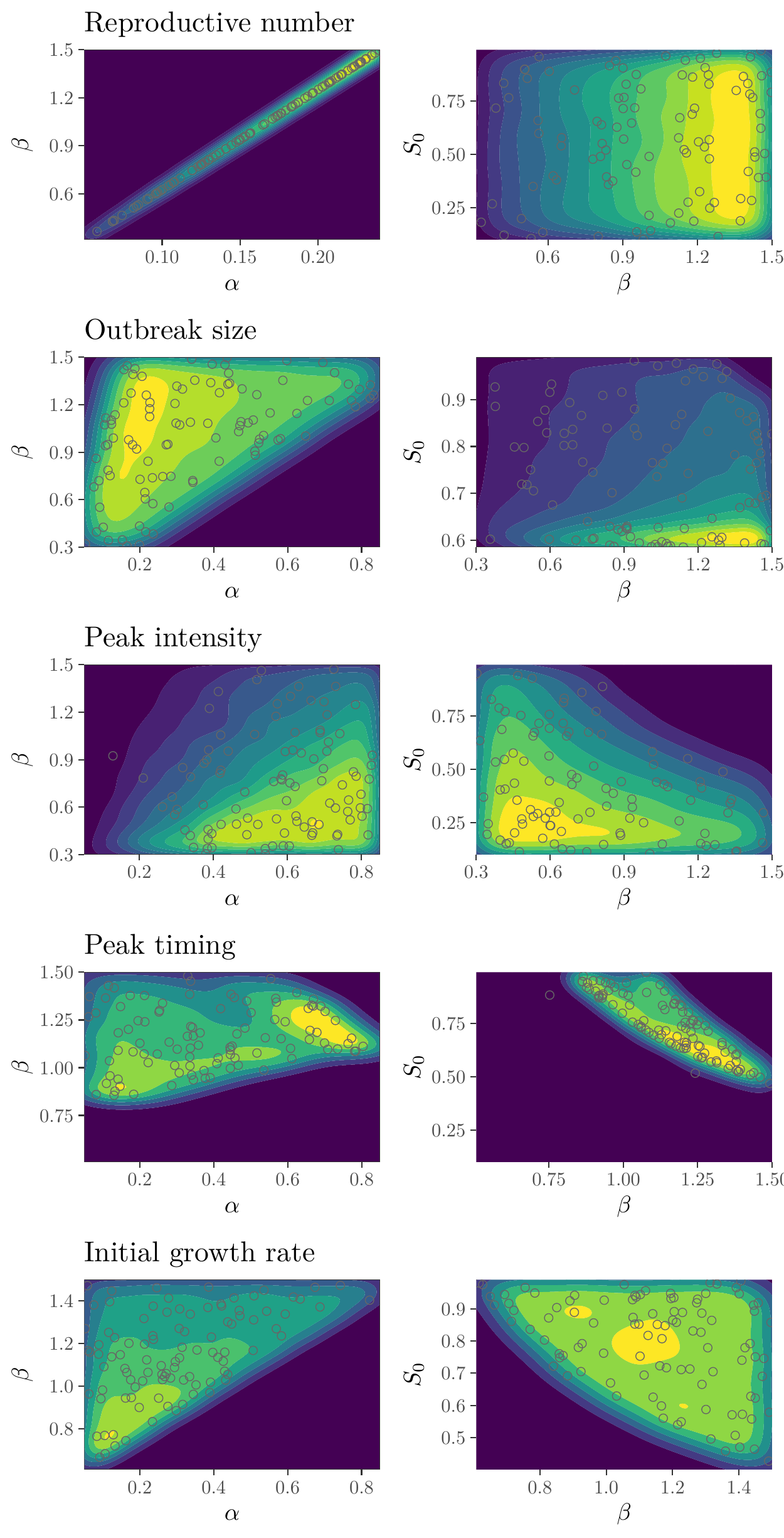}
    \caption{Density of 60,000 samples from $P(\alpha, \beta\mid u^*)$, and $P(\beta, S_0\mid u^*)$ given different summary transformations. True values and priors are the same as in the main text.}
    \label{fig:samples-cond-f}
\end{figure*}

Simulating from $P(\params\mid u^*)$ can be accomplished with the following procedure: let $\theta_i\in\params$ be a chosen ``pivot'' parameter and define $f^{-1}$ such that $f^{-1}(u, \params_{-i}) = \theta_i$. Then, assuming $\params$ are independent and that $f$ is bijective, using a change of variables and Bayes' rule we have
\begin{equation}
    P(\params_{-i}\mid u^*) \propto \prod_{j\neq i}P(\theta_j) \left|\frac{df^{-1}(u^*, \params_{-i})}{du}\right| P_{\theta_i}(f^{-1}(u^*, \params_{-i})).
    \label{eq:pdf-cond-f}
\end{equation}
Because $f$ is a deterministic function given $\params_{-i}$, samples from $P(\params\mid u^*)$ can therefore be obtained by first sampling $\params_{-i}^{(1)}, \dots, \params_{-i}^{(n)}$ from \eqref{eq:pdf-cond-f} using a standard simulation technique such as Accept-Reject sampling, and then letting $\theta_i^{(j)} = f^{-1}(u^*, \params_{-i}^{(j)})$. The result of sampling with this process for the five variables in Table \ref{tab:epi-vars} is shown in Figure \ref{fig:samples-cond-f}.

For example, under the transformation $f(\params) = \frac{\beta}{\alpha} =: \Rnot$, we define $f^{-1}(\alpha, S_0, \Rnot) = \alpha \Rnot$ and obtain
\begin{equation}
    P(\alpha, S_0 \mid \Rnot) \propto P_\alpha(\alpha)P_{S_0}(S_0) \alpha P_{\beta}\left(\Rnot\alpha\right).
    \label{eq:joint-given-Rnot}
\end{equation}
So we may sample $(\alpha^{(1)}, S_0^{(1)}), (\alpha^{(2)}, S_0^{(2)}), \dots$ from \eqref{eq:joint-given-Rnot}, then let $\beta^{(i)} = \Rnot^* \alpha^{(i)}$ to obtain a sample from $P(\alpha, \beta, S_0 \mid \Rnot)$.

For the final outbreak size, we define $\mathcal{O} := R(\infty) - R(0)$ to be the total proportion of individuals who end up in the recovered compartment due to infection. For $R(\infty)$ we have from \cite{Weiss2013},

\begin{equation}
    R(\infty) = 1 - S_0 \exp\left(-\Rnot (R(\infty) - R(0))\right),
\end{equation}
which we may use to solve for $\beta$ and obtain the inverse function
\begin{equation}
    \beta = \frac{-\alpha}{\osize} \log{\frac{1 - R(0) - \osize}{S_0}}
\end{equation}
and derivative
\begin{equation}
    \frac{d\beta}{d\osize} = \frac{\alpha}{\osize} \left(\frac{1}{\osize} \log{\frac{1 - R(0) - \osize}{S_0}} + \frac{1}{1 - R(0) - \osize}\right).
\end{equation}

For the peak intensity $\imax := \max_t I(t)$, to obtain samples from \eqref{eq:pdf-cond-f} we may use the equation
\begin{equation}
    \imax = I_0 + S_0 - \frac{\alpha}{\beta} \log S_0 - \frac{\alpha}{\beta} \left(1 + \log \frac{\alpha}{\beta}\right).
    \label{eq:imax}
\end{equation}
Although \eqref{eq:imax} yields only implicit solutions for any $\theta_i$, a closed-form solution for $S_0$ given $\imax$ can be found using Lambert's W,
\begin{equation}
    S_0 = -\Rnot^{-1} W_{-1}(-B),
\end{equation}
where $B = \exp\left(-\Rnot(\imax - I_0) - 1\right)$, and derivative
\begin{equation}
    \frac{d S_0}{d \imax} = \frac{1}{1 - B e^{W_{-1}(-B)}}.
\end{equation}

Derivation of the necessary equations for the initial growth rate $\grate := \beta S_0 = \alpha$ is straightforward. 

Finally, the peak timing $\tpeak$ does not have a known closed-form solution. Though more time-consuming, we can still approximate \eqref{eq:pdf-cond-f} by using univariate constrained optimization to evaluate the unknown $f^{-1}$, and adjoint methods to obtain the corresponding derivative.

% \bibliography{bibliography, bib-extra}

\end{document}